\documentstyle[aps,prl,multicol]{revtex}

\def\beginwide{
        \end{multicols} \vspace*{-0.5cm} \noindent
        \rule{3.5in}{.1mm}\rule{.1mm}{5mm} \widetext \medskip }
\def\beginwidetop{
        \end{multicols} \vspace*{-0.5cm} \noindent
        \widetext \medskip }
\def\endwide{
        \hspace*{3.5in}~\rule[-5mm]{.1mm}{5mm}\rule{3.5in}{.1mm}
        \begin{multicols}{2} \vspace*{-1.0cm} \noindent }
\def\endwidebottom{
        \begin{multicols}{2} \vspace*{-1.0cm} \noindent }

\draft

\begin{document}

\title{Self-organized criticality in linear interface depinning
and sandpile models}

\author{Alexei V\'azquez and Oscar Sotolongo-Costa}

\address{Department of Theoretical Physics, Faculty of
Physics, Havana University, Havana 10400, Cuba}

\maketitle

\begin{abstract}

The dynamics of an elastic interface profile $h(x,t)$ under a
driving force increasing at rate $c$, a restored force
$-\epsilon h$, and disorder is investigated. Using perturbation
theory and functional renormalization group the phase diagram
and the scaling exponents, up to the first order in
$\varepsilon=4-d$, are obtained. The model is found to be
critical in the double limit $\epsilon\rightarrow0$ and
$c/\epsilon\rightarrow0$ and belongs to a different universality
class as that of constant force models. It is shown that
undirected sandpile models with stochastic rules and linear
interface models with extremal dynamics belong to this new
universality class.

\end{abstract}


\begin{multicols}{2}

The problem of interface roughening in the present of quenched
disorder is a topic of recent interest, due to its importance as
a paradigm in condensed matter physics and due to the broad
range of applications. In general a $d$-dimensional self-affine
interface, described by a single-valued function $h(x,t)$,
evolves in a $(d+1)$-dimensional medium. Usually some form of
disorder affects the motion of the interface leading to its
roughening.  Earlier studies \cite{kardar} focuses on
time-independent uncorrelated disorder but most recent studies
analyses the motion of interfaces under quenched disorder
\cite{martys1,martys2,ji}. In the present of quenched disorder
an constant force driving two universality classes has been
found \cite{amaral}. One is described by the Kardar-Parisi-Zhang
equation \cite{kardar} with quenched noise. In this case the
interface is pinned by paths on a directed percolation cluster
of pinning sites \cite{tang1}. The second class is described by
the Edwards-Wilkinson (EW) equation \cite{edwards} with quenched
noise, linear interface depinning (LID).

The interface may be driven either by extremal dynamics
\cite{maslov,roux,paczuski1} or by constant force
\cite{leschhorn,dong,makse0}. While constant force models has
been extensively studied in the literature, either by functional
renormalization group (FRG) \cite{review,fisher} or numerical
simulations \cite{review}, extremal dynamics models are less
known.  However, in the last years extremal dynamics models have
gain more attention due to its relation with the theory of
self-organized criticality (SOC).

SOC was introduced to explain the critical behavior of a vast
class of driven dissipative system which evolve into a critical
state \cite{bak}. In its early state it was believed that such a
critical state is insensitive to changes in control parameters
and no fine-tuning is needed. More recent interpretations of
this phenomena have shown that criticality in SOC systems is
obtained after some control parameters, for instance the driving
and dissipation rates, are fine-tuned to zero \cite{vespignani}.

In the present work we focus our attention in the motion of a
$d$-dimensional interface profile obeying the following
equation,
\begin{equation} 
\lambda \partial_t h=\Gamma\nabla^2h+F-\epsilon h+\eta(x,h). 
\label{eq:0} 
\end{equation} 
with a force increasing at constant rate, $F=ct$. Here $\lambda$
is a friction coefficient, $\Gamma$ is the surface tension, $c$
is the driving rate, and $\epsilon$ a nonnegative constant. The
random force $\eta(x,h)$ is Gaussian distributed with zero mean
and
\begin{equation}
\langle\eta(x,h)\eta(x^{\prime},h^{\prime})\rangle =
\delta^d(x-x^{\prime})\Delta(h-h^{\prime}), 
\label{eq:1} 
\end{equation}
where $\Delta(h)$ is a monotonically decreasing function.

The aim of this work is to show that models described by this
equation are in a different universality class as constant force
models, and that undirected sandpile models with stochastic
rules and LID models with extremal dynamics belongs to this new
universality class. The motivation for the analogy between
sandpile models and LID models is based in a work by Paczusky
and Boettcher \cite{paczuski2}, where it is shown that a one
dimensional critical slope sandpile model is in the same
universality class as the depinning transition of a $d+1$
interface dragged at one end. Although this work pointed out the
analogy between LID and sandpile models its analysis was limited
to one dimension. We onclude that the existence of "spontaneous"
criticality in LID with extremal dynamics, as in sandpile model
\cite{vespignani}, is just a consequence of the unprecise
definition of these models.

If the force $F$ is constant and $\epsilon=0$ then eq.
(\ref{eq:0}) is reduced to the EW equation with quenched noise.
This case has been extensively studied in the literature
\cite{review,fisher}. A depinning transition takes place at
certain critical field force $F_c$ determined by the disorder.
For $F<F_c$ the interface is pinned after certain finite time
while for $F>F_c$ it moves with finite average velocity which
scales as $v\sim(F-F_c)^\beta$.

When $\epsilon>0$ and the force increases at rate $c$ then the
interface is never pinned by disorder, but always moves with a
finite average velocity $v$. A perturbative solution of eq.
(\ref{eq:0}) can thus be found expanding $h(x,t)$ around the
flat co-moving interface $vt$. Taking $h(x,t)=vt+y(x,t)$ we
obtain the following equation for $y(x,t)$
\begin{equation} 
\lambda \partial_t y=\Gamma\nabla^2y+(c-\epsilon v)t- 
\epsilon h -\lambda v +\eta(x,vt+h). 
\label{eq:2} 
\end{equation}

The average velocity is obtained using the constraint $\langle
y(x,t)\rangle=0$. For this purpose is better to work with the
equation for the Fourier transform $\tilde{h}(k$,$\omega)$. The
effective external field $(c-\epsilon v)t$ gives a singular term
of the order of $\omega^{-2}$. This singular term predominates
in the low frequency limit resulting, after imposing
$\langle\tilde{y}(k,\omega)\rangle=0$,
\begin{equation}
v=\frac{c}{\epsilon},
\label{eq:4}
\end{equation}
which is valid to all orders of perturbation expansion.

Another exact result can be obtained if one computes the
low-frequency and long-wavelength susceptibility. Adding a
source term $\varphi(x,t)$ to the right hand side of eq.
(\ref{eq:2}) and going to the Fourier space one obtains the
generalized response function
\begin{equation}
\tilde{G}(k,\omega)=\left\langle\frac{\tilde{h}(k,\omega)}
{\tilde{\varphi}(k,\omega)}\bigg|_{\tilde{\varphi}=0}\right\rangle=
\frac{1}{[\tilde{G}_0(k,\omega)]^{-1}-
\tilde{\Sigma}(k,\omega)},
\label{eq:6}
\end{equation}
where
\begin{equation}
[\tilde{G}_0(k,\omega)]^{-1}=\Gamma k^2-
i\lambda\omega+\epsilon
\label{eq:7}
\end{equation}
is the bare correlator and $\tilde{\Sigma}(k,\omega)$ is the
"self-energy". Since $\tilde{\Sigma}(0,0)=0$ and
$\tilde{G}_0(0,0)^{-1}=\epsilon$ it results that the
low-frequency and long-wavelength susceptibility (or simply the
susceptibility) is given by
\begin{equation}
\chi=\tilde{G}(0,0)=\epsilon^{-\gamma},\ \ \ \ \gamma=1.
\label{eq:8}
\end{equation}
This result is also exact to all orders of perturbation
expansion. Thus, when $\epsilon\rightarrow0$ the susceptivility
diverges and, therefore, the system is critical.

To go further we perform a FRG analysis of the problem,
following the general ideas developed for the constant force
case \cite{fisher}. We construct the generating functional
$Z=\int Dh D\hat{h}\exp(S)$ with action
\begin{equation} 
S=\int d^dxdt i\hat{h}[\lambda \partial_t
h-\Gamma\nabla^2h-F+\epsilon h-\eta(x,h)],
\label{eq:9} 
\end{equation} 
where $\hat{h}(x,t)$ is an auxiliary field. After averaging over
disorder an expansion around the mean-field (MF) solution yields
a generating functional with the low frequency form
\cite{fisher} $\bar{Z}\int Dy D\hat{y}\exp(\bar{S})$, where the
effective action $\bar{S}$ is given by
\beginwide
\begin{eqnarray}
\bar{S}=-\int d^dx dt\{[F-F_{\text{MF}}(v)]\hat{y}(x,t)
+\hat{y}(x,t)(\lambda\partial_t-\Gamma\nabla^2+\epsilon)y(x,t)\}
\nonumber\\
+\frac{1}{2}\int d^dx dt_1 dt_2 \hat{y}(x,t_1)y(x,t_2)
C[v(t_1-t_2)+y(x,t_1)-y(x,t_2)],
\label{eq:10}
\end{eqnarray}
\endwide
where $y$ and $\hat{y}$ are coarse grained versions of $h-vt$
and $-i\hat{h}$, respectively, and $C(h)$ is the MF correlation
function. Two differences appear with the constant force case.
First, in our case $F=ct$ and $F_{\text{MF}}(v)$, the MF force
corresponding to a velocity $v$, is given by
$F_{\text{MF}}(v)=c_{\text{MF}}(v)t$. Moreover, we have obtained
that $c=v\epsilon$ (see eq. (\ref{eq:4})) exactly, within and
beyond the MF approximation. Hence $F-F_{\text{MF}}(v)=0$.
Second, in the Gaussian part of the effective action there is an
extra term associated with the restored force, characterized by
the coefficient $\epsilon$. As it was shown above the
susceptibility diverges when $\epsilon\rightarrow0$. Thus,
$\epsilon$ is the control parameter of the interface described
by eq. (\ref{eq:0}). On the contrary in the constant force case
$\epsilon=0$ and $F-F_{\text{MF}}(v)$ is the control parameter.

The RG transformations are carried out as follows. We integrate
out the degrees of freedom in a momentum shell near the cutoff
$\Lambda$ and rescale $x\rightarrow bx$, $t\rightarrow b^zt$,
$y\rightarrow b^\zeta y$, and $\hat{y}\rightarrow
b^{\theta-d}\hat{y}$, where $b=\text{e}^l$ with $l\rightarrow0$.
As usual the cutoff $\Lambda$ appears because we start our
analysis from a coarse-grained equation, where we cannot resolve
spatial details smaller than $\Lambda^{-1}$.

The renormalization of the $\epsilon$ term yields
\begin{equation}
\frac{d\epsilon}{dl}=(\theta+z+\zeta)\epsilon,
\label{eq:11}
\end{equation}
which implies that the correlation length scale as
$\xi\sim\epsilon^{-\nu}$ with $\nu=1/(\theta+z+\zeta)$. Since
$\theta+z+\zeta)=2$ \cite{fisher} we finally obtain $\nu=1/2$,
which differs from the one obtained in the constant force case,
due to the existence of different control parameters. On the
contrary other scaling exponents results identical. For instance
\cite{fisher}
\begin{equation}
\zeta=\frac{\varepsilon}{3},\ \ \ \ z=2-\frac{2}{9}\varepsilon.
\label{eq:13}
\end{equation}
On the other hand, $vt$ and $y$ must scale in the same, so that
$v\rightarrow b^{\zeta-z}v$ yielding
\begin{equation}
\frac{dv}{dl}=(\zeta-z)v.
\label{eq:14}
\end{equation}
Thus, to reach the critical state both $\epsilon$ and $v$ should
be fine-tunned to zero, i.e. criticality is obtained in the
double limit $\epsilon\rightarrow0$ and
$v=c/\epsilon\rightarrow0$. From eq. (\ref{eq:14}) we define the
characteristic velocity $v_c\sim\epsilon^\beta$ with
\begin{equation}
\beta=\nu(z-\zeta).
\label{eq:15}
\end{equation}
Note that in constant force LID the average interface velocity
in the supercritical regime is given by $v\sim(F-F_c)^\beta$. On
the contrary, in the present model the average interface
velocity is fixed through eq. (\ref{eq:4}) and the exponent
$\beta$ just characterizes the $\epsilon$ dependence of the
characteristic velocity $v_c$, which delimit the subcritical and
supercritical regimes. For $v\gg v_c$ the noise term in eq.
(\ref{eq:2})is approximately annealed obtaining the EW equation
with annealed noise. In this case $\zeta=0$ for $d\geq2$. The
driving field predominates over disorder and, therefore, the
model is supercritical.

In the subcritical regime $\epsilon>0$ and $v\ll v_c$ the
dynamics takes place in the form of avalanches, characterized by
the avalanche size distribution $P(s)=s^{-\tau}g(s/s_c)$, where
$s_c\sim\epsilon^{-1/\sigma}$ is a cutoff avalanche size. In the
critical state $s_c\sim L^D$, where $D$ is the avalanche
dimension, and $\xi\sim L$ leading to the scaling relation
$\sigma=1/D\nu$.  Another scaling relation is obtained taking
into account that $\chi=\langle s\rangle$, leading to
$\gamma=(2-\tau)/\sigma$. On the other hand, for $d<d_c$, the
avalanche dimension and the roughness exponent are related via
$D=d+\zeta$ \cite{paczuski1}.  Using these scaling relations and
the values for $\gamma$, $\nu$, $\zeta$, and $z$ computed above
we obtain
\begin{equation}
D=d+\zeta,\ \ \ \
\tau=2(1-D^{-1}).
\label{eq:28}
\end{equation}

To investigate the analogy with sandpile models let us analyze a
discretized variant of eq. (\ref{eq:0}). In the cellular
automaton version of eq. (\ref{eq:0}) one defines the total
force
\begin{equation}
F_i=\sum_{nn}H_j-2dH_i+ht-\epsilon H_i+\eta_i(H_i),
\label{eq:18}
\end{equation}
and sites where $F_i>0$ are updated in parallel, advancing the
interface by one $H_i\rightarrow H_i+1$. Here
$\sum_{nn}H_j-2dH_i$ is a discretized Laplacian, where the sum
runs over nearest neighbors. Instead of follow the evolution of
the interface profile $H_i$ one may keep track of the total
force $F_i$. The evolution rules for the total force $F_i$ are
given by:  1- on each step $F_i$ is increased by $h$ in all
sites and 2- all sites where $F_i>0$ are updated in parallel
according to the toppling rule
\begin{equation} 
\begin{array}{ll} 
F_i\rightarrow F_i-2d-\epsilon+\eta_{ai}\\
F_{nn}\rightarrow F_{nn}+1,  
\end{array} 
\label{eq:19} 
\end{equation} 
where $\eta_{ai}=\eta_i(H_i+1)-\eta_i(H_i)$ is a zero mean
uncorrelated annealed noise and $nn$ denotes nearest neighbors.

From a simple inspection of eq. (\ref{eq:19}) one can see that
the total force follows the evolution rules of an undirected
sandpile automaton with an annealed noise, under a driving field
$h$ and with dissipation rate per toppling event $\epsilon$.
This class of sandpile models has been studied by Vespignani and
Zappery \cite{vespignani}, using mean-field and field theories.
They have obtained that $\gamma=1$ and $\nu=1/2$ are exact in
all dimensions. Moreover, their field theory predicts an upper
critical dimension $d_c=4$. These results are in agreement with
those obtained here and strongly suggest that both models are in
the same universality class.

In the mean-field and field theory by Vespignani and Zappery
conservation (understanding conservation as the balance between
input and output energy) is a necessary condition for
stationarity, which implies that the density of toppling
(active) sites is given by $\rho_a=h/\epsilon$. The equivalent
condition in our approach is found in eq. (\ref{eq:4}), which is
a consequence of the balance between the driving force $ht$ and
the average restored force $\epsilon vt$. The balance of these
two forces is thus our necessary condition for stationarity.
Moreover, since the interface advances only at active sites
(those where $F_i>0$), in one unit, then $v=\rho_a$, making the
connection between our approach and that of Vespignani and
Zapperi. Is also important to note that $v\leq1$, which is
consistent with the fact $\rho_a\leq1$.

Now, we proceed to show that the extremal LID model corresponds
to the critical state of eq. (\ref{eq:0}), i.e.
$h,\epsilon\rightarrow0$ and $v\rightarrow0$. The condition
$h\rightarrow0$ carry as a consequence that, if at time step
$t_0$ there are no active sites ($F_i<0$ for all sites) then at
time step $t_1=-F_j(t_0)/h$ the site $j$, with the maximum total
force, will be active. In the language of interface depinning
the interface at site $j$ will advance one unit, in the sandpile
language the site $j$ will topple. It is thus clear that if no
active site is present the system will follows extremal
dynamics. Now, what happened during the evolution of an
avalanche?

The active site $j$ will transfer energy to its nearest
neighbors, which at the same time may become active, and so on,
an avalanche is generated.  It is thus possible that at certain
time step $t$ there will more that one active site. These sites
will be updated in parallel according to the evolution rules
described above.  However, the order in which these sites are
updated is not important.  The process of toppling can never
transform any active site, different from itself, in inactive
and, therefore, the others active sites will remain active. On
the other hand, the energy transferred to its neighbors is
constant, independent of the total force at this site. Moreover,
the site with the maximum total force will be always among the
set of active sites. Hence, one can arrange the sequence of
toppling events, of actives sites at time step $t$, in such a
way that always the site with the maximum total force topples.

Finally, in the double limit $h\rightarrow0$,
$\epsilon\rightarrow0$ the total force is reduced to the local
force
\begin{equation}
F_i\rightarrow f_i=\sum_{nn}H_j-2dH_i+\eta_i(H_i),
\label{eq:29}
\end{equation}
Hence, in the critical state of the cellular automaton version
of eq.  (\ref{eq:0}) one may arrange the sequence of toppling
events in such a way that the site which topples has the maximum
local force $f_i$, which is the extremal dynamics variant of
LID. We thus conclude that extremal dynamics models corresponds
to the critical state of the LID model described by eq.
(\ref{eq:0}).

In table \ref{tab:1} some numerical estimates for the LID model
with extremal dynamics and the Manna $d$-state model, the
prototype of stochastic sandpile model, are given.  Our FRG
estimates from eqs. (\ref{eq:13}) and (\ref{eq:28}) are also
shown for comparison. Increasing the lattice dimension the FRG
predictions get closer to the numerical estimates, obtaining a
complete agreement in three dimensions. Using the numerical
estimates for $\tau$ and $D$, reported for the LID model with
extremal dynamics, we have tested the scaling relation
$(2-\tau)D=\gamma\nu^{-1}=2$, within the numerical error it is
fulfilled. In all cases the difference in the numerical
estimates for both models are contained in the error bars, which
implies that they are in the same universality class. Our
predictions are thus confirmed with the numerical simulations.

In the discussion we have not included the Bak-Tang-Wiesenfeld
(BTW) model neither a sandpile model with stochastic dissipation
introduced by Chessa {\em et al} \cite{chessa}. In the first
case because the BTW model is deterministic and it is not clear
yet if stochastic and deterministic models are in the same
universality class. The LID model introduced here was mapped
into a sandpile model with annealed noise and, therefore, our
conclusions cannot be extended to deterministic models. In the
second case because according to the simulations by Chessa {\em
et al} \cite{chessa} the upper critical dimension for their
stochastic model is 6, in disagreement with $d_c=4$ as obtained
from our FRG analysis and the numerical simulations of the
$d$-state Manna model \cite{lubeck}. On the theoretical side,
$d_c=4$ is in agreement with previous reports by
D\'{\i}az-Guilera, using a dynamic renormalization group
approach for the BTW and Zhang models \cite{guilera}, and by
Vespignani {\em et al}
\cite{vespignani1}.

In summary, we have shown that the existence of "spontaneous"
criticality in LID with extremal dynamics, as in sandpile model
\cite{vespignani}, is just a consequence of the unprecise
definition of these models. SOC corresponds to the onset of
nonlocality in the dynamics of the interface.  Nonlocality, and
hence criticality, is obtained by fine tuning the control
parameters, precisely as in continuous phase transitions. The
extremal dynamics corresponds with a fine tuned interface
depinning transition at constant velocity. It was also
demonstrated that LID with extremal dynamics and undirected
sandpile models with stochastic rules belong to the same
universality class, which is different from that of constant
force LID.

We thanks A. Vespignani for helpful comments and suggestions.
This work was partially supported by the {\em Alma Mater} prize,
given by the University of Havana.

%%%%%%%%%%% tables

\begin{table}\narrowtext
\begin{tabular}{lllllll}
$d$ & Model & $\tau$ & $z$ & $D$ & $(2-\tau)D$ & Ref.\\ \hline
1 & LID & 1.13(2) & & 2.23(3) & 1.94(7) & \cite{paczuski1}\\
  & FRG & 1 & $\frac{4}{3}\approx1.33$  & 2 & 2\\
\\
2 & LID & 1.29(2) & & 2.75(20) & 1.95(7) & \cite{paczuski1}\\
  & Manna & 1.273 & 1.500 & 2.750 & 2.00 &  \cite{lubeck}\\
  & FRG & $\frac{5}{4}=1.25$ & $\frac{14}{9}\approx1.56$ & 
  $\frac{8}{3}\approx2.67$ & 2\\
\\
3 & LID & & & 3.34(1)$^{*}$ & & \cite{review}\\
  & Manna & 1.40 & 1.75 & 3.33 & 2.00 & \cite{lubeck}\\
  & FRG & $\frac{7}{5}=1.4$ & $\frac{16}{9}\approx1.78$ & 
  $\frac{10}{3}\approx3.33$ & 2
\end{tabular}
\caption{Scaling exponents for the LID model with extremal
dynamics and the Manna $d$-state model. Results obtained here
using FRG are shown for comparison. $^{*}$ It was computed using
the scaling relation $D=d+\zeta$ and the reported value of
$\zeta$.}
\label{tab:1}
\end{table}

\end{multicols}

\end{document}